# Photon echo from localized excitons in semiconductor nanostructures

S.V. Poltavtsev[a,b,], I.A. Yugova[b], I.A. Akimov[a,c], D.R. Yakovlev[a,c], M. Bayer[a,c]

[a] *Experimentelle Physik 2, Technische Universität Dortmund*
*D-44227, Dortmund, Germany*

[b] *Spin optics laboratory, St. Petersburg State University,*
*198504, St. Petersburg, Peterhof, Russia*

[c] *Ioffe Institute, Russian Academy of Sciences,*
*194021, St. Petersburg, Russia*

**Abstract**

An overview on photon echo spectroscopy under resonant excitation of the exciton complexes in semiconductor nanostructures is presented. The use of four-wave-mixing technique with the pulsed excitation and heterodyne detection allowed us to measure the coherent response of the system with the picosecond time resolution. It is shown that, for resonant selective pulsed excitation of the localized exciton complexes, the coherent signal is represented by the photon echoes due to the inhomogeneous broadening of the optical transitions. In case of resonant excitation of the trions or donor-bound excitons, the Zeeman splitting of the resident electron ground state levels under the applied transverse magnetic field results in quantum beats of photon echo amplitude at the Larmor precession frequency. Application of magnetic field makes it possible to transfer coherently the optical excitation into the spin ensemble of the resident electrons and to observe a long-lived photon echo signal. The described technique can be used as a high-resolution spectroscopy of the energy splittings in the ground state of the system. Next, we consider the Rabi oscillations and their damping under excitation with intensive optical pulses for the excitons complexes with a different degree of localization. It is shown that damping of the echo signal with increase of the excitation pulse intensity is strongly manifested for excitons, while on trions and donor-bound excitons this effect is substantially weaker.

e-mail: ilja.akimov@tu-dortmund.de

## 1. Introduction

Coherent optical spectroscopy provides rich information about the energy structure and the natural linewidth of the spectral lines of the studied system. Use of pulsed light sources makes it possible to conduct experiments with the temporal resolution and get the information on the dynamical processes and the main mechanisms which lead to loss of coherence (phase relaxation). The inherent property of the majority of macroscopic systems is the inhomogeneous broadening of the optical transitions. Resonant excitation of such a system by a train of optical pulses results in a photon echo phenomenon – a nonlinear coherent response of the system in a form of delayed optical pulse [1,2]. Delay times, at which the photon echo can be observed, are defined by the coherent properties of the local (individual) quantum-mechanical excited state. Thereby, the study of photon echoes allows one to overcome the inhomogeneous broadening of optical transitions and acquire the information about the coherent dynamics of the individual excitation in a large ensemble of emitters. Two-pulse and three-pulse photon echoes are actively exploited to investigate the energy structure and coherent evolution of the optical excitations in atomic systems, rare-ion crystals and semiconductors [3-5]. Moreover, photon echo is considered as a possible candidate for the realization of the optical memory based on the ensemble of emitters [6].

In semiconductors, the elementary optical excitations are exciton complexes (coupled electron-hole pairs). Excitons possess a large oscillator strength, which permits the rapid and efficient optical excitation using the sub-picosecond laser pulses. It should be noted that the exciton-exciton interaction in semiconductor crystals results in a complex dynamics of the optical coherent response [4]. Such dynamics can be observed on a time scale of several picoseconds. In case of strong exciton localization, however, many-body interactions are suppressed and localized excitons can be considered as single non-interacting complexes with a discrete energy spectrum. In the simplest case of the resonant excitation, exciton complex can be treated as a two-level energy system with single ground and excited states. Fluctuations in the composition and the localization potential determine the inhomogeneous broadening of the optical transitions, which leads to the formation of photon echoes [7,8]. A good example of the system with the localized excitons is the ensemble of self-organized quantum dots. Four-wave-mixing (FWM) experiments with the use of femtosecond laser pulses have shown that, at low temperature, the exciton coherence time $T_2$ can be comparable with its lifetime $T_1 \sim 1$ ns [9]. Thus, the localization of the exciton results in an extension of its coherence time. Moreover, since many-body interactions are suppressed, there is a possibility of coherent manipulation of exciton states using the intense laser pulses. As an example of such coherent manipulation, the Rabi oscillations in two-level system can be demonstrated [10,11].

Special interest represent the systems with more than two electron states interacting with light. Here, the most remarkable examples are the V- (Λ-) type energy systems, where the single ground (excited) state is optically coupled with the two excited (ground) states [12]. Such systems allow observation of different interesting phenomena, including quantum beats, coherent population trapping and electromagnetically-induced transparency [6,13]. The main feature of the Λ-type scheme is the large coherence time of the ground states. In semiconductors, Λ-type scheme is realized when localized resident carriers, e.g. electrons in the conduction band or holes in the valence band, are optically excited and their spin degree of freedom is involved. In case of resident electrons, the negatively charged exciton (trion, $X^-$) or the donor-bound exciton ($D^0X$) are possible optical excited states [14]. Optical manipulation of the resident carrier spin using the short picosecond laser pulses has been demonstrated in different semiconductor systems [15-17]. However, the majority of the FWM and photon echo studies in semiconductors did not use so far the spin degree of freedom in the ground state [18-21].

In this paper, we present an overview of recent results on photon echo spectroscopy using the resonant excitation of the localized exciton complexes. A special attention is devoted to the systems with the resident electrons under the application of transverse magnetic field. In this case, the step-like stimulated Raman process following after the excitation of the system by the two optical pulses allows the coherent transfer of the optical excitation into the spin ensemble of the resident electrons and observation of the long-lived photon echoes [22,23]. The long-lived photon echo monitors the local dynamics of the ground state of the system and provides rich information about the spin processes in the resident electron ensemble even if the optical transitions are strongly broadened [24]. The choice of material and dimensionality of the quantum system plays a crucial role here. They influence the selection rules of the optical transitions, the localization of exciton complexes and the spin relaxation processes for both excited and ground states. In the paper the main effects are considered for a model system – CdTe/(Cd,Mg)Te quantum wells (QWs), where inhomogeneous broadening of the optical transitions is small and it is possible to excite selectively the various exciton complexes with the different degree of localization [24,25]. We also consider the photon echo signals in the hexagonal epitaxial ZnO layers [26], ZnSe/(Zn,Mg)(S,Se) QWs and self-organized (In,Ga)As/GaAs quantum dots (QDs) placed inside the planar microcavities in order to enhance the exciton-light coupling. The experimental results are obtained by means of the four-wave-mixing with heterodyne detection described in section **2**. In section **3**, long-lived photon echoes measured from the ensemble of resident electrons are considered. In section **4**, the Rabi oscillations at the intensive optical excitation, which is the one of the necessary conditions to perform an efficient optical control of quantum states, are described.

**2. Experimental technique**

In order to study FWM and photon echo signals the experimental set-up featuring high sensitivity and picosecond temporal resolution was developed. The optical scheme of the set-up is shown in Fig. 1. As a source of picosecond laser pulses the Ti:sapphire laser Mira-900 tunable in the range 700-1000 nm and pumped by the 532 nm Verdi-V10 laser was exploited. Laser pulses with the duration of about 2 ps are split by the non-polarizing beamsplitters into the first, second, and third excitation pulses as well as the reference pulse. All pulses, except the first one, are delayed by means of the mechanical translation stages, as shown in Fig. 1: the second pulse is delayed by the interval $\tau_{12}$ with respect to the first one, the third pulse – by the interval $\tau_{23}$ with respect to the second one, and the reference pulse – by the interval $\tau_{Ref}$ with respect to the first one. The sample is placed into the helium bath cryostat and cooled down to the temperature of about 2 K. Three excitation pulses are focused to the sample into the spot of about 300 μm in diameter using the spherical mirror with the focal length of 500 mm. All laser pulses are linearly copolarized. The first and the second pulses with the wavevectors $k_1$ and $k_2$ hit the sample under 3° and 4°, accordingly. The third pulse propagates in the same direction as the second one ($k_3 = k_2$). The FWM signal is collected in the reflection geometry in $2k_2 - k_1$ direction by means of the same spherical mirror. The signal is split equally into two channels of the balanced photodetector using the non-polarizing beamsplitter. The reference pulse is guided to the same photodetector avoiding the sample and the cross-correlation between the reference pulse and FWM signal is measured.

In order to detect weak FWM signal the technique of optical heterodyning is applied. For that purpose, the two acousto-optic modulators AOM-1 and AOM-2 are used to shift the optical frequencies of the first excitation pulse and the reference pulse by –81 MHz and +80 MHz, respectively. The optical frequency of the FWM signal amounts to $\nu_{FWM} = 2\nu_2 - \nu_1$, where $\nu_1 = \nu_0 - 81$ MHz, $\nu_2 = \nu_0$ ($\nu_0$ is the original frequency of light), which results in $\nu_{FWM} = \nu_0 + 81$ MHz. The interference of FWM signal with the reference pulse field ($\nu_{Ref} = \nu_0 + 80$ MHz) gives rise to the optical beats at the differential frequency $\Delta f = 1$ MHz, which are measured at the photodetector output. The DC components, which are defined by the total intensities of the detected beams, are canceled in the balanced detection, while the amplitude of optical beats is doubled. The latter is detected by a "fast" lock-in at the frequency of 1 MHz and its magnitude is proportional to the modulus of the product of the reference pulse amplitude $E_{Ref}$ and the FWM signal amplitude $E_{PE}$: $\delta I_{Det} \sim |E_{PE} E_{Ref}^*|$. In order to exclude the spurious signals we modulate the intensity of the first excitation pulse using a chopper at the frequency of about 1 kHz. The second stage of the detection is performed at this frequency by means of a "slow" lock-in. External magnetic field up to 6 Tesla is applied along the sample plane. In experiments with the wide-band-gap semiconductor nanostructures (ZnO, ZnSe) the second harmonic generation (SHG) unit is additionally used in order to convert the original infrared (IR) ps-pulses into the ultraviolet (UV) pulses with the wavelength of 350 – 495 nm and duration of 1.3 ps. FWM signal from the ZnO-based samples with the polished substrate is detected in the transmission geometry using another 500-mm spherical mirror and the same detection scheme.

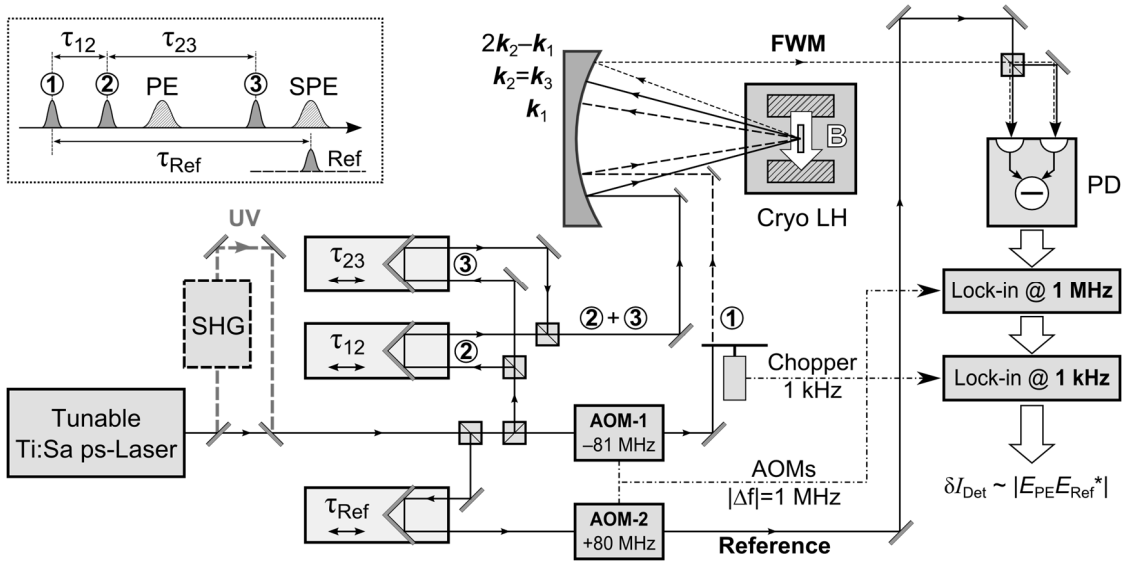

Fig. 1. Schematic presentation of the experimental set-up for measuring the photon echoes with the picosecond time resolution. Notations: SHG – second harmonic generation unit; Cryo LH – helium bath cryostat, PD – photodetector, AOM – acousto-optic modulator; UV – ultraviolet laser light. Inset: temporal diagram of the excitation pulses and echo signals. Two-pulse echo (PE) and three-pulse echo (SPE) are delayed by $\tau_{12}$ after the second and the third pulse, accordingly.

The studied objects are various $A_2B_6$ and $A_3B_5$ epitaxial heterostructures including CdTe/(Cd,Mg)Te QWs [22-24], ZnSe/(Zn,Mg)(S,Se) QWs, (In,Ga)As/GaAs QDs as well as ZnO epitaxial layers [26]. The resonances in QWs and bulk materials possess large oscillator strength, which does not cause problems with the detection of FWM signal. The oscillator strength of the QD excitons is weaker. Moreover, the strong inhomogeneous broadening significantly limits the amount of QDs, whose transition energy matches the resonant optical excitation. In order to enhance the FWM signal QDs are incorporated inside the microcavity, which comprises, for example, two Bragg mirrors formed during the growth of the structure. Use of microcavities with small Q-factor (100-200) ensures that the photon mode transmits the picosecond laser pulse without any distortion and the effects of strong coupling do not take place.

The described set-up allows for measurement of different contributions to the FWM signal, in particular, two-pulse photon echo (PE) and three-pulse PE (stimulated PE, SPE) propagating in time in accordance with the diagram shown in the inset of Fig. 1. Variation of the reference pulse delay $\tau_{Ref}$ makes it possible to measure the PE temporal profile with the picosecond resolution. Figure 2(a) displays the temporal profiles of the PEs measured in four different systems: donor-bound exciton ($D^0X$) in the CdTe/(Cd,Mg)Te single QW, $D^0X$ in the 140 nm-thick ZnO epitaxial layer, trion in ZnSe/(Zn,Mg)(S,Se) QW, as well as excitons in (In,Ga)As/GaAs QDs placed in the Bragg microcavity. Since we detect cross-correlation between the FWM signal and the reference pulse, the PE profile corresponds to the convolution of the echo pulse with the reference pulse. The width of PE profile reflects partly the inhomogeneous broadening of the excited ensemble, which is related to the reversal phase relaxation rate: $\Gamma_2^* \propto 1/T_2^*$. When the spectral width of the ensemble is significantly narrower than the laser pulse spectrum, the temporal PE profile is accordingly longer than the laser pulse duration. PE from the $D^0X$ complex in CdTe/(Cd,Mg)Te QW with the duration of about 8 ps corresponds to this case. However, when the inhomogeneous broadening of the ensemble is comparable or substantially broader than the laser pulse spectrum, the echo pulse profile approximately corresponds to the pulse auto-correlation function. This is the case for PE measured from the other systems in Fig. 2(a). It should be mentioned that the UV laser pulse becomes $\sqrt{2}$ times shorter after the conversion from the IR range. The PE profile, however, may also have a complicated shape, when the excitation intensity is sufficiently increased. In this case, measurements of the FWM signal go beyond the $\chi^{(3)}$ regime and enter the regime of Rabi oscillations. Section **4** of this paper is devoted to Rabi oscillations measurements.

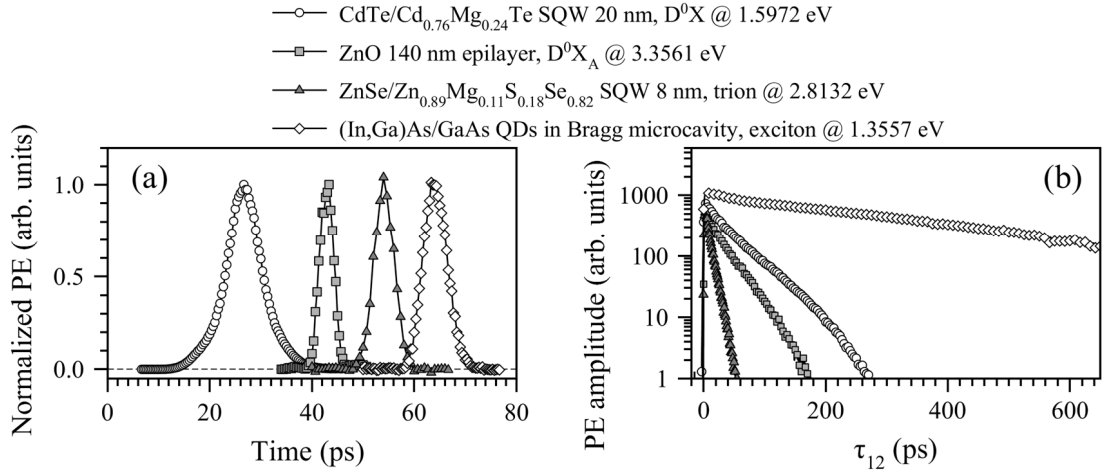

Fig. 2. The PE temporal profile (a) and PE amplitude decay (b) measured from $D^0X$ in CdTe/(Cd,Mg)Te single QW (circles), $D^0X$ in ZnO epitaxial layer (squares), trion in ZnSe/(Zn,Mg)(S,Se) QW (triangles), and exciton in (In,Ga)As/GaAs QDs placed in the Bragg microcavity (diamonds).

When the second pulse delay is varied simultaneously with the reference pulse delay under the condition $\tau_{Ref} = 2\tau_{12}$, then PE amplitude decay can be measured. Figure 2(b) shows the PE decay dynamics for the four studied systems. The decays can be approximated with the single exponential function $\sim\exp(-2\tau_{12}/T_2)$, from which the time of irreversible phase relaxation (coherence time) $T_2$ of the studied resonance is obtained. It corresponds to the homogeneous spectral linewidth (full width at half-maximum) $\Gamma_2 = 2\hbar/T_2$ of the resonance. The longest time $T_2 = 750$ ps is measured on the excitons in (In,Ga)As/GaAs QDs ensemble, which corresponds to $\Gamma_2 = 1.8$ μeV. The shortest time $T_2 = 16$ ps is evaluated for the trion in ZnSe/(Zn,Mg)(S,Se) QW, for which $\Gamma_2 = 82$ μeV. The reversible phase relaxation time $T_2^*$ and irreversible phase relaxation time $T_2$ do not correlate in general case and can be measured independently using the photon echo technique.

By tuning the wavelength of the excitation pulses one can measure the spectral dependence of the PE decay dynamics. Spectral dependences of $T_2$ obtained in that way on the CdTe/(Cd,Mg)Te QW and ZnO epitaxial layer are shown in Fig. 3. Interestingly, in both systems we observe non-monotonous dependence of $T_2$ in the vicinity of $D^0X$ resonances where coherence time decreases with increasing the photon energy. This observation contradicts to the common hypothesis about longer coherence times of the exciton complexes with the lowest energy of the optical transition due to the stronger localization. Apparently, it is not applied in case of the donor-bound excitons in the studied systems.

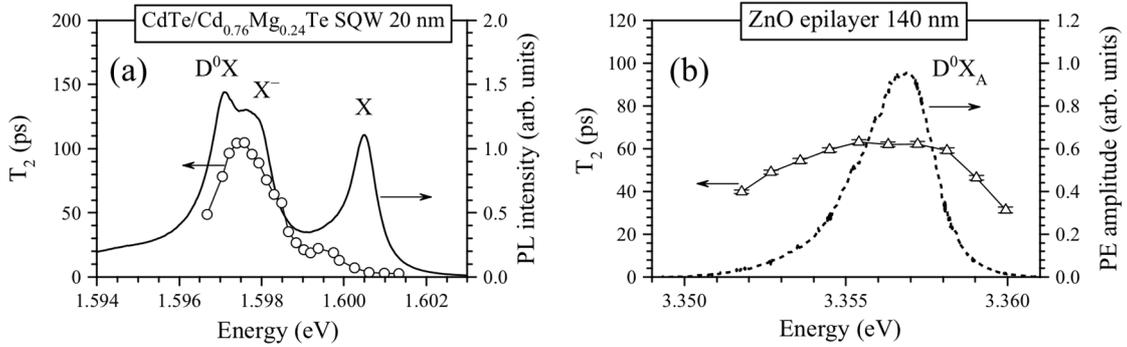

Fig. 3. Spectral dependences of irreversible phase relaxation time $T_2$ measured in CdTe/(Cd,Mg)Te single QW (a) and in 140 nm-thick ZnO epitaxial layer (b). Solid line in panel (a) corresponds to the PL spectrum with the following resonances: exciton (X), trion ($X^-$) and donor-bound exciton ($D^0X$). Dashed line in panel (b) is the PE amplitude spectrum measured at $\tau_{12} = 27$ ps.

## 3. Long-lived photon echo from ensemble of resident electrons

This section describes the studies in the weak excitation regime $\chi^{(3)}$, when the FWM signal intensity depends linearly on the intensity of every excitation pulse (pulse energy ~ 10 – 100 nJ/cm$^2$).

Let us consider PE signal from the trions localized in CdTe/(Cd,Mg)Te QW. The energy level structure and optical transitions are shown in Fig. 4(a). The ground state corresponds to the resident electron and is defined by the doublet with the electron spin $S = 1/2$. Optically excited state with the lowest energy corresponds to the trion with the zero electron spin (singlet state). Thus, the trion angular momentum $J = 3/2$ is defined by the heavy hole and the excited state represents a doublet as well. Under the transverse magnetic field B applied along the QW plane (perpendicular to the structure growth axis), the electron spin states are split by the energy $\hbar\omega_L = g\mu_B B$, where $\omega_L$ – is the Larmor precession frequency, g – is the electron g-factor and $\mu_B$ – is the Bohr magneton. The trion doublet splitting is not large because of the strong heavy hole g-factor anisotropy in the QW structures [29]. The resonant excitation of the trion is performed by the light pulses propagating along the structure growth axis, i.e. perpendicular to the magnetic field axis (Voigt geometry). In this case, the optical transitions are allowed between all four states, which are defined by the angular momentum projection to the magnetic field direction. These transitions are linearly polarized along (H) or perpendicular (V) to the magnetic field axis, as shown in Fig. 4(a). The similar level scheme and selection rules work for the donor-bound excitons. Thereby, our consideration is applied not only to the trion, but also to the $D^0X$ complex. Moreover, the similar energy level scheme in the transverse magnetic field and selection rules for the optical transitions are realized for $D^0X$ complex in the hexagonal ZnO bulk crystal, if the optical excitation is performed along the *c*-axis of the crystal. In that case, the crystal field leads to the splitting of the hole states in the valence band and plays the same role as the confinement potential in a quantum well.

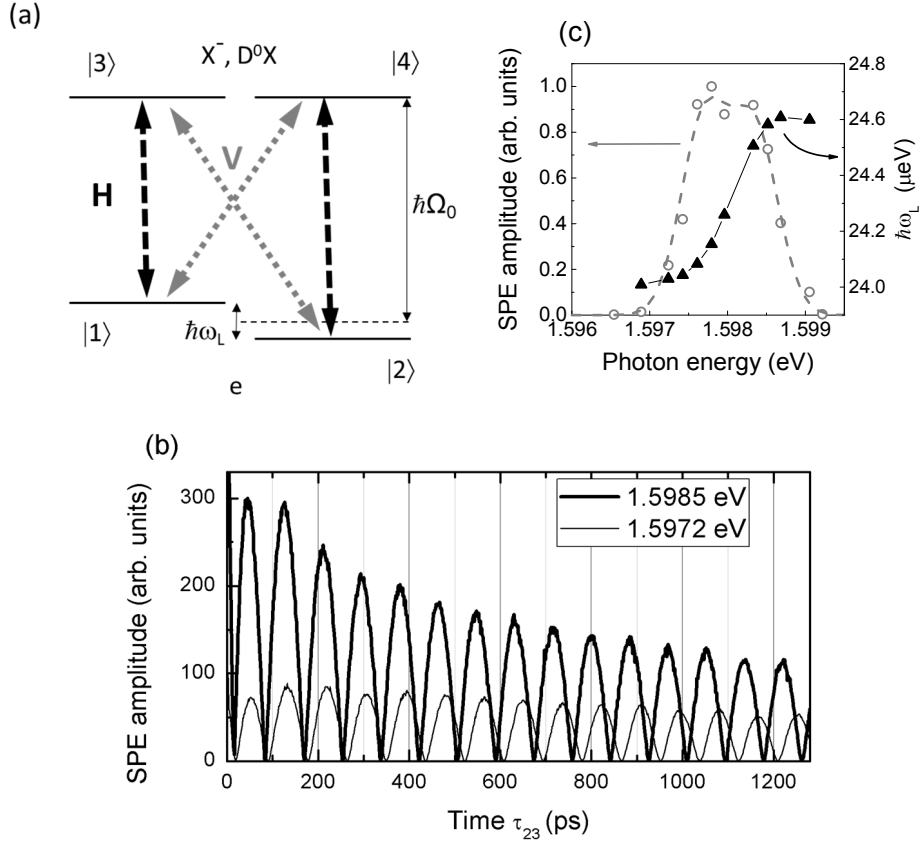

Fig. 4. (a) Scheme of energy levels and optical transitions describing the excitation process of the trion ($X^-$) or donor-bound exciton ($D^0X$) in the singlet state in the CdTe/(Cd,Mg)Te QW. H and V correspond to the linear polarizations of the optical transitions parallel or perpendicular to the magnetic field axis, respectively. (b) SPE amplitude as a function of $\tau_{23}$ delay time for the resonant excitation of the trion (1.5985 eV) and $D^0X$ (1.5972 eV) complexes measured at T = 2 K, B = 260 mT, $\tau_{12}$ = 27 ps. Polarization sequence of the excitation pulses is HVV. (c) Spectral dependences of the SPE amplitude (circles) and the Zeeman splitting in the ground state $\hbar\omega_L$ (triangles).

For the demonstration of the long-lived PE we accomplish the optical excitation of the system by application of a train of three laser pulses. Proper choice of the light polarization in the pulse train provides an additional selectivity between the different excitation paths [24]. We consider polarization sequence HVV (the first pulse polarized along H, the second and third pulses polarized along V). This corresponds to the most interesting case when all optical transitions are exploited and the coherent superposition between the one pair of states is transferred into the other pair of states after the each excitation event in a step-like process.

In this case, the first pulse (H-polarized) results in the excitation of trions by means of the optical transitions between the states $|1\rangle$ and $|3\rangle$ at the frequency $\Omega_0 - \omega_L/2$ or between the states $|2\rangle$ and $|4\rangle$ at the frequency $\Omega_0 + \omega_L/2$. Here, $\Omega_0$ corresponds to the trion resonant frequency in the absence of magnetic field ($\omega_L = 0$). The first pulse creates coherent superpositions between the states pairs $|1\rangle$-$|3\rangle$ and $|2\rangle$-$|4\rangle$, i.e. the optical polarization is generated. Using the density matrix notation, this polarization corresponds to the nondiagonal elements $\rho_{13}$ and $\rho_{24}$. It is implied here that, before the first pulse arrival, the system is in the ground state with the zero spin polarization, i.e. the Zeeman splitting of the levels $\hbar\omega_L$ is small compared to the Boltzmann energy $k_B T$ and the only nonzero density matrix elements are $\rho_{11} = \rho_{22} = 1/2$. Here, $k_B$ – is the Boltzmann constant, and T – is the crystal temperature. This condition is well fulfilled at the small magnetic fields up to 1 Tesla at T = 2 K.

Let us assume that coherence of every excited trion is preserved until the second pulse arrival, despite the fact that the macroscopic polarization of the medium rapidly decays due to the inhomogeneity of the optical transitions (reversal dephasing). The second pulse (V-polarized) stimulates the optical transition down into the ground state in such a way that the optical coherences $\rho_{13}$ and $\rho_{24}$ are transferred into the spin coherence of the electron ensemble $\rho_{12}$. In this state, the optical dephasing process is frozen and the further evolution of the system is determined only by the spin dynamics of the resident electrons under the applied external magnetic field. It should be stressed that the electron spin relaxation time can exceed the trion lifetime by several orders of magnitude [30]. The third laser pulse (also V-polarized) again optically excites the trions, thus, creating the optical polarization $\rho_{42}$ and $\rho_{31}$. This initiates the rephasing process and results in emission of the long-lived SPE pulse.

In the case, when the Zeeman splitting of the electron levels in the ground state is smaller than the spectral width of the laser pulse (~1 meV) and the inhomogeneous broadening of the optical transitions, the SPE signal is well described by the Gaussian-shape pulses with the amplitude

$$P \propto e^{-\frac{2\tau_{12}}{T_2}} \left[ e^{-\frac{\tau_{23}}{\tau_T}} \cos(\omega_L \tau_{12}) + e^{-\frac{\tau_{23}}{T_2^e}} \cos(\omega_L (\tau_{12} + \tau_{23})) \right], \qquad (1)$$

where $\tau_T$ – is the trion spin lifetime. Equation (1) is the solution of the Lindblad equation in the short rectangular laser pulse approximation [23,24]. In addition it is assumed that the pulse duration is small as compared to the Larmor spin precession period in the ground state $T_L = 2\pi/\omega_L$. In turn, $T_L$ is considered to be shorter than the second pulse delay $\tau_{12}$. Equation (1) comprises two terms. The first term on the right-hand side is responsible for the spin relaxation and the trion recombination (up to 100 ps in CdTe/(Cd,Mg)Te QWs), which decays rapidly. The second term, in turn, is determined by the time of transverse spin relaxation in the ensemble of localized resident electrons $T_2^e$. It is the latter term which is responsible for the long-lived SPE signal occurring when $T_2^e \gg \tau_T$.

From Eq.(1) it follows that the three-pulse SPE amplitude under the applied magnetic field oscillates at the Larmor precession frequency. Therefore, it is possible to measure the Zeeman splitting in the ground state using the selective resonant optical excitation. It should be noted that the measurement is possible even when the homogeneous spectral linewidth of the optical transition $2\hbar/T_2$ is larger than the energy splitting since the long-lived SPE signal is defined exclusively by the broadening of the ground state levels. Thereby, the method described can be exploited as an optical coherent spectroscopy with the high spectral resolution.

The dynamics of the long-lived SPE under the resonant excitation of the trions and donor-bound excitons in CdTe/(Cd,Mg)Te QW is presented in Fig. 4(b). The time interval between the second and the third pulses $\tau_{23}$ was varied and the SPE signal was measured at the temporal position $\tau_{Ref} = 2\tau_{12} + \tau_{23}$. This experiment is performed at the external magnetic field strength B = 260 mT and the second pulse delay $\tau_{12} = 27$ ps. From Fig. 4(b) it follows that the variation of time interval $\tau_{23}$ results in an oscillating signal decaying on the time scale of several nanoseconds, which is significantly longer than the optical coherence time $T_2$ measured in the same sample as shown for the different excitation energies in Fig. 3(a). We stress that the long-lived SPE signal is observed only in the presence of the magnetic field in the spectral range of 1.597-1.599 eV, i.e. for resonant excitation of the trion and $D^0X$ complexes. It is clear that for observation of the long-lived SPE signals it is necessary to address the optical transitions with resident electrons in the ground state.

Using Eq. (1) we retrieved the spectral dependence of long-lived SPE amplitude as well as the oscillation frequency $\omega_L$ and the decay time $T_2^e$. Spectral dependence of the Zeeman splitting at B = 260 mT is shown in Fig. 4(c). The most remarkable feature here is the variation of the oscillation frequency with the photon energy demonstrating the step-like behavior. Namely, $\hbar\omega_L$ increases from 24.0 up to 24.6 μeV when the excitation energy is tuned from the spectral position of $D^0X$ to that of the trion. From these data we evaluated g-factors $|g| = 1.595$ and 1.635 for the resident electrons bound to donors and electrons localized on the potential fluctuations, respectively. Thereby, for the case of CdTe/(Cd,Mg)Te QW we have measured the splitting between the Zeeman sub-levels of the electrons in the ground state with the accuracy below 1 μeV. This revealed the difference between the two electron sub-ensembles: The donor-bound electrons and electrons localized on the potential fluctuations.

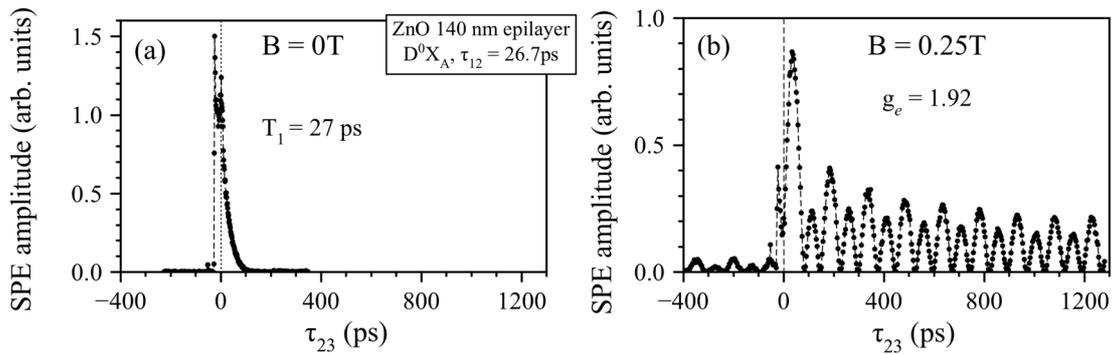

Fig. 5. Dependence of the SPE amplitude on $\tau_{23}$ delay time for the resonant excitation of $D^0X_A$ complex (photon energy 3.3566 eV) in the 140 nm-thick ZnO epitaxial layer measured at T = 2 K and $\tau_{12} = 27$ ps. Polarization sequence of the excitation pulses is HVV. (a) B = 0 T; (b) B = 250 mT.

It should be noted that long-lived SPE signal, whose dynamics is dictated by the spin coherence in the ground state, can be observed also in other semiconductor systems. An example is the ZnO bulk crystal under the resonant excitation of the A donor-bound excitons ($D^0X_A$ complex). Measurements performed on the 140 nm-thick epitaxial ZnO layer are demonstrated in Fig. 5. In this case the hexagonal crystal axis (*c*-axis) is directed perpendicular to the sample plane and, therefore, practically coincides with the propagation direction of the excitation pulses. In the absence of the magnetic field, the SPE signal decays exponentially with the lifetime of the $D^0X_A$ complex, i.e. SPE amplitude $\sim\exp(-\tau_{23}/T_1)$. Here, the decay time is $T_1 = 27$ ps [26]. Application of the magnetic field results in the oscillating long-lived SPE signal. From the oscillation period $T_L=150$ ps at B=250 mT we deduce |g| = 1.92, which corresponds to the electron g-factor. We note that the spin dephasing time $T_{2e}$ in ZnO is longer than that in the CdTe/(Cd,Mg)Te QW, which is apparently determined by smaller scattering of g-factor values in the ensemble of the donor-bound electrons. From Fig. 5(b) it follows that oscillating signal is observed even at the "negative" $\tau_{23}$ delays. This delay range corresponds to the large $\tau_{23}$ time intervals comparable with the laser pulse repetition period (13 ns). In that case, an interesting regime may be realized, in which, similar to the effect of resonant spin amplification [31], accumulation of the spin polarization under excitation with a periodic sequence of optical pulses would lead to the amplification or damping of the SPE signal depending on the magnetic field amplitude.

## 4. Rabi oscillations

Resonant excitation of the optical transitions with sufficiently large intensity makes it possible to switch from $\chi^{(3)}$ regime of photon echo generation to the regime of coherent Rabi oscillations, which might be observed in certain systems [25,27,28]. Since the generation of the echo signal requires at least two optical pulses, the amplitude of each of them can be varied for that purpose. In order to perform such experiments the setup was equipped with controllable optical attenuators in both excitation beams. The measurement procedure includes consistent increase of the amplitude for one of the excitation pulses after each acquisition of two-pulse PE temporal profile. This results in a two-dimensional presentation of Rabi oscillations in PE amplitude. The measurements were performed in zero magnetic field where the energy scheme of optical transitions can be reduced to the two-level system with single ground and excited states.

Figure 6 displays the experimental data and the theoretical simulations obtained for the different optical transitions in the CdTe/(Cd,Mg)Te single QW [25]. It is seen that the measurements of Rabi oscillations are very sensitive to the excitation energy which can be compared with the photoluminescence spectrum, shown in Fig. 3(a). Excitation of the localized exciton in the energy range 1.5985 – 1.5990 eV does not show oscillations in PE amplitude when the first pulse amplitude is increased, but it causes a strong damping of the echo amplitude [see Fig. 6(a)]. This damping is due to the excitation-induced dephasing (EID) processes emerging due to weak exciton localization [4].

In the case of the localized trion the situation changes drastically. Figures 6 (b) and (c) show the results of the photon echo transients when scanning the amplitude of the first and second pulses, respectively. In the first case, we observe the oscillations in PE amplitude, where the second maximum is shifted in the time, and PE amplitude is damped due to the EID. In the second case, no temporal shift of PE is observed and the oscillations are strongly smoothed.

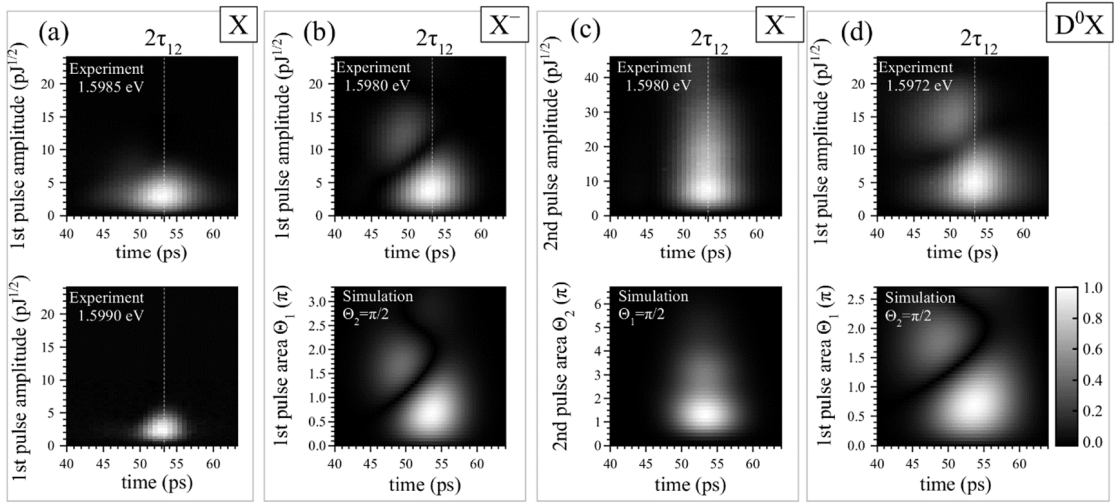

Fig. 6. Rabi oscillations in the PE amplitude measured in the CdTe/(Cd,Mg)Te single QW: (a) Variation of the first pulse amplitude for excitation of the exciton at two different energies; (b) and (c) variation of the first and the second pulse amplitudes, respectively, for excitation of the trion; (d) variation of the first pulse amplitude for excitation of the $D^0X$ complex. Upper panels of (b)-(d) are experimental data; bottom panels are theoretical simulations. The excitation pulse areas are measured in units of $\pi$.

The temporal shift of PE amplitude with variation of the first pulse amplitude were observed also in the ensemble of (In,Ga)As/GaAs QDs embedded in the Bragg microcavity [27]. This effect occurs when the spectrally broad ensemble is excited by the spectrally narrow laser pulses. As a result, the ensemble experiences dephasing already during the action of optical pulse due to the large scattering of the resonant frequencies of the oscillators with respect to the central frequency of light. When the second pulse reverses the temporal evolution of the ensemble, the rephasing of oscillators is slightly shifted in time and PE is advanced with respect to the $2\tau_{12}$ time [32]. It is possible to describe this effect in a simple model of two-level system ensemble, whose coherent dynamics follows the solutions of optical Bloch equations. The results of the numerical calculation describing the trion experimental data are shown in the bottom panels of Fig. 6(b) and 6(c). Apart from the EID process, the spatial inhomogeneity of the excitation spot approximated by the Gaussian shape is also taken into account [25].

Rabi oscillations detected from the $D^0X$ transition appear qualitatively similar to those measured from the trion, but there are few differences. First, the echo pulse duration observed from $D^0X$ complex is longer than that from the trion, which corresponds to the smaller inhomogeneous broadening of the $D^0X$ ensemble. And second, the EID influence is lower for the $D^0X$ complex than for the trion, which results in a slower decay of Rabi oscillations, when the excitation pulse amplitude is increasing.

The photon echo measurements demonstrate that the ensemble of the $D^0X$ complexes in the CdTe/(Cd,Mg)Te QWs with a small donor concentration ($\sim 10^{10}$ cm$^{-2}$) represents a good system for generation of efficient photon echo signal using the sequence of two laser pulses with the areas of $\pi/2$ and $\pi$, respectively.

However, further increase of the pulse power leads to unavoidable losses of coherence and attenuation of the photon echo signal due to many-body interactions and heating of the electron system.

**Acknowledgements**

The authors acknowledge their colleagues L. Langer, M. Salewski, T. Meier, M. Reichelt, M. M. Glazov, L. E. Golub, G. G. Kozlov and Yu. V. Kapitonov. The authors thank Deutsche Forschungsgemeinschaft ICRC TRR-160 (Project A3) and Russian Foundation for Basic Research (Project № 15-52-12016 NNIO_a) for the financial support. This work is also supported by the St-Petersburg University 11.34.2.2012 grant.